\documentstyle[12pt,epsf]{article}
\newcommand{\beq}{\begin{equation}}
\newcommand{\eeq}{\end{equation}}
\newcommand{\beqa}{\begin{eqnarray}}
\newcommand{\eeqa}{\end{eqnarray}}
\def\simgt{\rlap{\lower 3.5 pt \hbox{$\mathchar \sim$}} \raise 1pt \hbox {$>$}}
\def\simlt{\rlap{\lower 3.5 pt \hbox{$\mathchar \sim$}} \raise 1pt \hbox {$<$}}

\newcommand{\be}{\begin{equation}}
\newcommand{\ee}{\end{equation}}
\newcommand{\bea}{\begin{eqnarray}}
\newcommand{\eea}{\end{eqnarray}}

\catcode`\@=11
\def\@citex[#1]#2{\if@filesw\immediate\write\@auxout{\string\citation{#2}}\fi
  \def\@citea{}\@cite{\@for\@citeb:=#2\do
    {\@citea\def\@citea{,\penalty\@m}\@ifundefined
       {b@\@citeb}{{\bf ?}\@warning
       {Citation `\@citeb' on page \thepage \space undefined}}%
\hbox{\csname b@\@citeb\endcsname}}}{#1}}
\def\citer{\@ifnextchar [{\@tempswatrue\@citexr}{\@tempswafalse\@citexr[]}}
\def\@citexr[#1]#2{\if@filesw\immediate\write\@auxout{\string\citation{#2}}\fi
  \def\@citea{}\@cite{\@for\@citeb:=#2\do
    {\@citea\def\@citea{--\penalty\@m}\@ifundefined
       {b@\@citeb}{{\bf ?}\@warning
       {Citation `\@citeb' on page \thepage \space undefined}}%
\hbox{\csname b@\@citeb\endcsname}}}{#1}}
\begin{document}

\begin{titlepage}
\begin{flushright}
        CERN-TH/97-353\\
        hep-ph/9712302\\
        December 1997\\
\end{flushright}
\vskip 1.8cm
\begin{center}
 \boldmath
{\Large\bf Two-loop Correction to the Leptonic Decay\\[0.2cm]
of Quarkonium}
\unboldmath
\vskip 1.8cm
{\sc M. Beneke} \hspace{0.2cm}and \hspace{0.2cm} {\sc A. Signer}
\vskip .3cm
{\it Theory Division, CERN, CH-1211 Geneva 23}
\vskip .4cm
and
\vskip .4cm
{\sc V.A. Smirnov}
\vskip .3cm
{\em Nuclear Physics Institute, Moscow State University, \\
119889 Moscow, Russia}
\vskip 2.0cm
\end{center}

\begin{abstract}
\noindent Applying asymptotic expansions at threshold, 
we compute the two-loop QCD correction to the
short-distance coefficient that governs the leptonic
decay $\psi\to l^+ l^-$ of a $S$-wave quarkonium state and
discuss its impact on the relation between the quarkonium
non-relativistic wave function at the origin and the
quarkonium decay constant in full QCD.\\[1cm]
\end{abstract}

\vfill

\end{titlepage}


\noindent Quarkonium decays played an
important role in establishing Quantum
Chromodynamics (QCD) as a weakly interacting theory at short distances.
Calculations of heavy quarkonium decays usually proceed
under the assumption that the heavy quark-antiquark bound state is
non-relativistic and that the decay amplitude factorizes into
the bound state wave function at the origin and a short-distance
quark-antiquark annihilation amplitude. One can then explain the
small width of the $S$-wave spin-triplet charmonium state $J/\psi$,
because it can decay only through
electro-magnetic annihilation or annihilation into at least three
gluons \cite{APP75}. Today's understanding of quarkonium
bound states has refined
this picture and allows us to calculate relativistic corrections
systematically at the expense of introducing further non-perturbative
parameters that characterize the bound state. Such calculations
can be done most transparently in the framework of a non-relativistic
effective field theory (NRQCD) \cite{CL,BBL} that
implements the factorization of
contributions to the (partial) decay widths from different length scales.
Besides potentially large relativistic corrections, the size of
radiative corrections to the quark-antiquark annihilation amplitudes
has always been a matter of concern. The one-loop radiative corrections
to the decays $\psi\to l^+ l^-$ (where $l=e,\mu$) \cite{oneloop},
$\psi\to\,$light hadrons,
$\psi\to \gamma+\,$light hadrons \cite{radiative} are large and question
the practicability of factorization for charmonium and, perhaps,
even bottomonium. [Here and in the following we use $\psi$ as a label
for any $S$-wave spin-triplet state, i.e. $J/\psi$ or $\psi'$ for
charmonium and $\Upsilon(nS)$ for bottomonium.]

In this Letter we address this question and
report on the calculation of two-loop short-distance corrections to
the leptonic decay $\psi\to l^+ l^-$. This is the first (and also
`easiest') two-loop matching calculation for quarkonium decays
and it can also be used to connect the decay constant of the $\psi$ meson
defined in QCD with the non-relativistic wave function at the origin,
which appears in NRQCD and potential models. In turn,
this wave function is an important input parameter for the prediction
of other quarkonium decays and also quarkonium production
cross sections.

We recall that $\psi$ decays leptonically through interaction with the
electro-magnetic current. The partial decay rate, neglecting the tiny
lepton masses, is exactly given by
\begin{equation}
\Gamma(\psi\to l^+ l^-) = \frac{4\pi e_Q^2 \alpha_{em}^2 f_\psi^2}
{3 M_\psi},
\end{equation}
where $M_\psi$ is the mass of $\psi$, $\alpha_{em}$ the fine structure
constant and $e_Q$ the electric charge of the heavy quark in units of the
electron charge.
The $\psi$ decay constant $f_\psi$
is defined through the following matrix element of the
electro-magnetic current:
\begin{equation}
\label{decayconstant}
\langle \psi(p)|\bar{Q}\gamma_\mu Q|0\rangle = (-i) f_\psi M_\psi
\,\epsilon^*_\mu(p).
\end{equation}
[$\epsilon_\mu(p)$ is the $\psi$ polarization vector and $p$ the
$\psi$ momentum.] The leptonic
decay rates are known experimentally \cite{PDG}.
For $J/\psi$ we find $f_{J/\psi}=(405\pm 15)\,$MeV.

The decay constant parametrizes the strong interaction effects and
contains long- and short-distance
contributions. For quarkonium the short-distance scale is $1/M_\psi$
and the long-distance (bound state) scales are $1/(M_\psi v)$ and
$1/(M_\psi v^2)$, where $v$ is the (small) characteristic velocity
of the $\psi$'s quark constituents. The short-distance contributions
can be isolated, and calculated in perturbation theory, by matching
the vector current in QCD onto a series of operators in NRQCD. Up
to corrections of order $v^4$, the matching relation is given by
\begin{eqnarray}
\label{currents}
\langle \psi(p)|\bar{Q}\gamma^\mu Q|0\rangle &=&
\Lambda^{\mu i}(p) \Bigg[C_0\!\left(\alpha_s,\frac{m_Q}{\mu}\right)
\langle \psi|\psi^\dagger
\sigma_i\chi|0\rangle(\mu)
\nonumber\\
&&+ \,\frac{C_1\!\left(\alpha_s,\frac{m_Q}
{\mu}\right)}{6 m_Q^2}
\,\langle \psi|\psi^\dagger\vec{D}^2\sigma_i\chi|0\rangle(\mu)
+ {\cal O}(v^4) \Bigg],
\end{eqnarray}
where $\psi$ (not to be confused with the $\psi$ meson) and
$\chi$ denote non-relativistic two-spinors, $\vec{D}$ the spatial
covariant derivative and $m_Q$ the heavy quark mass. [More details on
notation and NRQCD can be found in Ref.~\cite{BBL}. Note, however, that
we use a relativistic normalization of states also for the
matrix elements in NRQCD.] The matrix elements on the right-hand side are
defined in the $\psi$ rest frame and $\Lambda(p)$ is the
matrix that performs the Lorentz boost into this frame.
The matching coefficients $C_0$ and $C_1$
are expressed as series in the strong coupling $\alpha_s$ and account
for the short-distance QCD corrections. The matching coefficients and
matrix elements in NRQCD individually depend on the factorization
scale $\mu$. $C_1$ is defined (as in Ref.~\cite{BC97})
such that $C_1=1+{\cal O}(\alpha_s)$ and \cite{oneloop}
\begin{equation}
\label{matching}
C_0\!\left(\alpha_s,\frac{m_Q}{\mu}\right) =
1-\frac{2 C_F \,\alpha_s(m_Q)}{\pi} + c_2\!\left(m_Q/\mu\right)
\left(\frac{\alpha_s}{\pi}\right)^2+\ldots,
\end{equation}
where $C_F=(N_c^2-1)/(2 N_c)$ and $N_c=3$ is the number of colours. We
now discuss the calculation of the two-loop matching coefficient $c_2$.

\begin{figure}[t]
   \vspace{-1.5cm}
   \epsfysize=19cm
   \epsfxsize=12cm
   \centerline{\epsffile{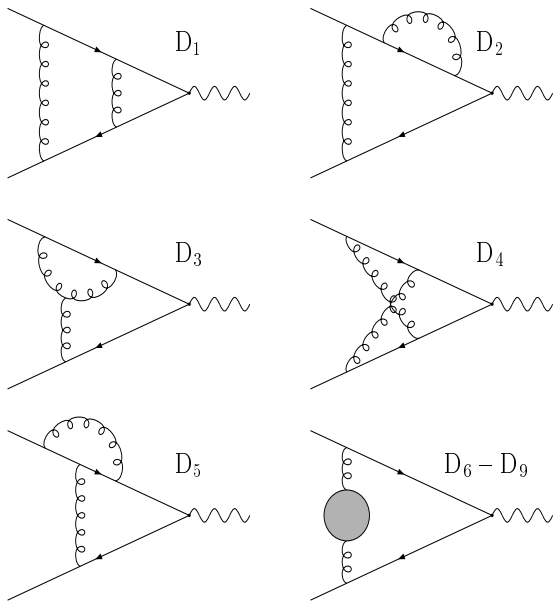}}
   \vspace*{-8.3cm}
\caption[dummy]{\small Diagrams that contribute to $\Gamma_{\rm QCD}$.
Symmetric diagrams exist for $D_{2,3,5}$. The last diagram summarizes
vacuum polarization contributions from massless fermions ($D_6$),
gluons ($D_7$), ghosts ($D_8$) and the massive fermion with mass
$m_Q$ ($D_9$).
\label{fig1}}
\end{figure}

Since the matching coefficient contains only short-distance effects,
it can be obtained by replacing the quarkonium state $\psi$ on both
sides of Eq.~(\ref{currents}) by a free quark-antiquark pair of
on-shell quarks at small relative velocity. In terms of this on-shell
matrix element, the matching equation can be rewritten as
\begin{equation}
\label{newmatch}
Z_{2,{\rm QCD}}\,\Gamma_{\rm QCD} =
C_0\,Z_{2,{\rm NRQCD}}\,Z_J^{-1}\,\Gamma_{\rm NRQCD} + {\cal O}(v^2),
\end{equation}
where $Z_2$ are the on-shell wave function renormalization constants
in QCD and NRQCD and $\Gamma$ the amputated, bare electro-magnetic
annihilation vertices in QCD and NRQCD. The Feynman diagrams
for $\Gamma_{\rm QCD}$
at two loops are shown in Fig.~\ref{fig1}. Since the current
$J=\psi^\dagger\sigma_i\chi$ need not be conserved in NRQCD,
we allowed for its renormalization, $J_{\rm bare}=Z_J J_{\rm ren}$,
on the right-hand side. We then obtain $C_0$ by calculating all
other quantities in Eq.~(\ref{newmatch}) in dimensional regularization
and using the modified minimal subtraction scheme
($\overline{\rm MS}$ scheme) \cite{msbar}.

The matching calculation is considerably simplified, if one uses
the threshold expansion of Ref.~\cite{BS} to compute $\Gamma_{\rm QCD}$
directly as an expansion in $v^2$. The threshold expansion is
obtained by writing down contributions corresponding to hard
($l\sim m_Q$), soft ($l\sim m_Q v$), potential ($l_0\sim m_Q v^2$,
$l_i\sim m_Q v$) and ultrasoft ($l\sim m_Q v^2$) regions. The contributions
from soft, potential and ultrasoft loop momenta can all be identified
with diagrams in NRQCD that appear in the calculation of
$\Gamma_{\rm NRQCD}$. Hence, they drop out of the matching relation
Eq.~(\ref{newmatch}) and it suffices to compute the
contribution to the threshold expansion of the diagrams in Fig.~\ref{fig1},
where all loop momenta are hard. [The threshold expansion is not
only convenient; it also provides an implicit definition of NRQCD
in dimensional regularization. This is necessary, because dimensionally
regularized NRQCD is not given by the dimensionally regularized Feynman
integrals constructed from the vertices and propagators of NRQCD.
In order to avoid that the cut-off for the effective theory be 
treated as larger than $m_Q$, dimensionally regularized NRQCD has to
be supplemented by a prescription for expanding the Feynman
{\em integrands}. This prescription is provided to all orders as part
of the diagrammatic threshold expansion method \cite{BS}.]

We briefly describe the calculation of the hard contributions
to $\Gamma_{\rm QCD}$. [Details of the calculational method
and a solution to the recurrence algorithm for the two-loop integrals
will be given in a long write-up of this Letter.] The spinor
structure of the on-shell
matrix element in QCD is conventionally parametrized by two form
factors, $F_1$ and $F_2$, of which only the combination $F_1+F_2$ is
required here. Since terms of order $v^2$ are not needed to determine
$C_0$ [see Eq.~(\ref{newmatch})], we may set the relative momentum to
zero and compute the form factors directly at threshold. The
form factors have Coulomb singularities at threshold and diverge as
$1/v^2$. However, these singularities appear only in the soft,
potential and ultrasoft
contributions, and the hard contribution is well-defined directly
at threshold in dimensional
regularization. The loop integrals simplify considerably, once the
relative momentum is set to zero, since they then depend only on a
single scale. We then project on the form factor $F_1+F_2$ and reduce
all integrals to integrals without numerators. These integrals
can be further reduced to `simple' integrals and two non-trivial two-loop
integrals by means of recurrence relations derived from integration
by parts in the loop momenta \cite{IBP}. The solution to the recurrence
relations is the difficult part of the calculation. The remaining
non-trivial two-loop integrals can be calculated explicitly using
standard Feynman parameters. The results obtained for the diagrams
of Fig.~\ref{fig1} are listed in Table~\ref{diagrams}.

\begin{table}[t]
\begin{center}
\begin{tabular}{|c|c|c|c|c|}\hline
\rule[-0.35cm]{0cm}{0.95cm}
 & Colour factor
 & $\frac{1}{\epsilon^2}$
 & $\frac{1}{\epsilon}$
 & finite\\
\hline
$D_1 \rule[-3mm]{0mm}{10mm}$ & $C_F^2$
& $\frac{9}{32}$
& $-\frac{27}{64}-\frac{5\pi^2}{24}$
& $-\frac{81}{128}-\frac{133\pi^2}{96}-\frac{5\pi^2\ln 2}{12}-
\frac{35\zeta(3)}{8}$ \\[0.15cm]
$D_2$ & $C_F^2$
& $-\frac{3}{16}$
& $-\frac{43}{32}$
& $\frac{733}{192}+\frac{971\pi^2}{576}$ \\[0.15cm]
$D_3$ & $C_F C_A$
& $\frac{15}{32}$
& $-\frac{5}{64}-\frac{\pi^2}{16}$
& $\frac{715}{384}-\frac{319\pi^2}{576}-\frac{\pi^2\ln 2}{8}-
\frac{21\zeta(3)}{16}$ \\[0.15cm]
$D_4$ & $C_F \,(C_A-2 C_F)$
& $0$
& $\frac{3}{16}-\frac{\pi^2}{16}$
& $-\frac{39}{32}-\frac{251\pi^2}{1152}-\frac{3\pi^2\ln 2}{8}-
\frac{31\zeta(3)}{16}$ \\[0.15cm]
$D_5$ & $C_F \,(C_A-2 C_F)$
& $-\frac{9}{32}$
& $-\frac{19}{64}$
& $\frac{761}{384}+\frac{1157\pi^2}{1152}+\frac{\pi^2\ln 2}{6}-
\frac{3\zeta(3)}{4}$ \\[0.15cm]
$D_6$ & $C_F T_F n_f$
& $-\frac{1}{8}$
& $\frac{5}{48}$
& $-\frac{355}{288}-\frac{5\pi^2}{48}$ \\[0.15cm]
$D_7$ & $C_F C_A$
& $\frac{19}{128}$
& $-\frac{53}{768}$
& $\frac{6787}{4608}+\frac{95\pi^2}{768}$ \\[0.15cm]
$D_8$ & $C_F C_A$
& $\frac{1}{128}$
& $\frac{1}{768}$
& $\frac{361}{4608}+\frac{5\pi^2}{768}$ \\[0.15cm]
$D_9$ & $C_F T_F$
& $-\frac{1}{4}$
& $\frac{13}{48}$
& $-\frac{145}{96}+\frac{5\pi^2}{72}$ \\[0.25cm]
\hline
\rule{0cm}{0.5cm}
Sum & $C_F^2$ & $\frac{21}{32}$ & $-\frac{99}{64}-\frac{\pi^2}{12}$
    & $\frac{637}{384}-\frac{733\pi^2}{576}+\zeta(3)$ \\[0.15cm]
    & $C_F C_A$ & $\frac{11}{32}$ & $-\frac{49}{192}-\frac{\pi^2}{8}$
    & $\frac{4811}{1152}+\frac{209\pi^2}{576}-\frac{\pi^2\ln 2}{3}-4\zeta(3)$
    \\[0.15cm]
    & $C_F T_F n_f$ & $-\frac{1}{8}$ & $\frac{5}{48}$
    & $-\frac{355}{288}-\frac{5\pi^2}{48}$ \\[0.15cm]
    & $C_F T_F$ & $-\frac{1}{4}$ & $\frac{13}{48}$
    & $-\frac{145}{96}+\frac{5}{72}$ \\[0.15cm]
\hline
\end{tabular}
\end{center}
\caption[dummy]{\small Coefficient of $(\alpha_s/\pi)^2\,(e^{\gamma_E}\,
m_Q^2/(4\pi\mu^2))^{-2\epsilon}$ for the hard contribution to
the diagrams of Fig.~\ref{fig1}
evaluated at threshold $q^2=4 m^2$. $D_{2,3,5}$ include a factor
of 2 to account for the corresponding symmetric diagrams. For $SU(3)$
the colour factors are $C_F=4/3$, $C_A=3$, $T_F=1/2$. The number of
light (massless) quark flavours is denoted by $n_f$. \label{diagrams}}
\end{table}

After summing all the diagrams, multiplying by the two-loop QCD on-shell
wave function renormalization constant \cite{Z2}, and performing
(one-loop) coupling and mass renormalization the result still contains
poles in $\epsilon=(4-d)/2$ (where $d$ is the space-time dimension).
Since the wave function renormalization constant $Z_{2,{\rm NRQCD}}$ in
NRQCD equals 1
up to higher-order terms in $v^2$, not needed here, these poles
are attributed to an anomalous dimension of $J$, which first arises
at the two-loop order. As a consequence the matrix element
$\langle \psi|\psi^\dagger\sigma_i\chi|0\rangle(\mu)$ is
factorization scale-dependent.
We define it in the $\overline{\rm MS}$ scheme and obtain
the anomalous dimension for the NRQCD vector current $J$:
\begin{equation}
\label{andim}
\gamma_J = \frac{\mbox{d}\ln Z_J}{\mbox{d}\ln\mu} =
-C_F\,(2 C_F+3 C_A)\,\frac{\pi^2}{6}\left(\frac{\alpha_s}{\pi}
\right)^2 + {\cal O}(\alpha_s^3).
\end{equation}
The scale dependence is compensated by the scale dependence of the
two-loop matching coefficient $c_2(m_Q/\mu)$ in Eq.~(\ref{matching}).
Separating the different colour group factors, the final result for
$c_2(m_Q/\mu)$ in the $\overline{\mbox{MS}}$ scheme is:
\begin{eqnarray}
c_2(m_Q/\mu) &=& C_F^2\,c_{2,A} + C_F C_A\,c_{2,NA} + C_F T_F\,n_f\,
c_{2,L} + C_F T_F\,c_{2,H}, \label{2loop} \\[0.1cm]
c_{2,A}  &=& \pi^2\left[\frac{1}{6}\,\ln\left(\frac{m_Q^2}{\mu^2}\right) -
\frac{79}{36}+\ln 2\right] + \frac{23}{8}-\frac{\zeta(3)}{2}, \label{c2A}\\
c_{2,NA} &=& \pi^2\left[\frac{1}{4}\,\ln\left(\frac{m_Q^2}{\mu^2}\right) +
\frac{89}{144}-\frac{5}{6}\,\ln 2\right] - \frac{151}{72}-
\frac{13\zeta(3)}{4},\\
c_{2,L}  &=& \frac{11}{18},\\
c_{2,H}  &=& -\frac{2\pi^2}{9}+\frac{22}{9}. \label{c2H}
\end{eqnarray}
Here $\zeta(3)=1.202\ldots$ and we have taken all
fermions with masses less than $m_Q$ as massless,
which is a good approximation even for $m_Q=m_b$, the bottom quark mass,
in which case we neglect
$m_c$, the charm quark mass. The coefficient $c_{2,L}$
proportional to $n_f$, the number of light fermions,
has been recently obtained by Braaten and Chen \cite{BC97}.
Note also that the $C_F^2$-term
of the form factors $F_{1,2}$ close to threshold has been
calculated by Hoang \cite{hoang}, using the absorptive parts of the
form factors obtained in Ref.~\cite{absorptive}. Because the result in
Ref.~\cite{hoang} contains hard and soft (potential, ultrasoft)
contributions, it is not possible to extract the matching
coefficient $c_{2,A}$ from Ref.~\cite{hoang}. [The structures in
$c_{2,A}$ that cannot arise from the small loop momentum
regions agree with Hoang's result.]

The size of the two-loop correction to $C_0$ [Eq.~(\ref{matching})]
given by Eqs.~(\ref{2loop})--(\ref{c2H}) is enormous. We define the
(scale-dependent) non-relativistic decay constant as
\begin{equation}
\langle \psi|\psi^\dagger\sigma_i\chi|0\rangle(\mu) = (-i)
f_\psi^{\rm NR}(\mu) M_\psi\,\epsilon^*_i,
\end{equation}
in analogy with Eq.~(\ref{decayconstant}). The non-relativistic
decay constant is related to the
$\psi$ wave function at the origin by $M_\psi\,(f_\psi^{\rm NR})^2
= 12\,|\Psi(0)|^2$. Using Eq.~(\ref{currents}), we obtain
\begin{equation}
\label{rel}
f_\psi = \left(1-\frac{8\alpha_s(m_Q)}{3\pi}-(44.55-0.41 n_f)
\left(\frac{\alpha_s}{\pi}\right)^2+\ldots\,\right)
f_\psi^{\rm NR}(m_Q).
\end{equation}
With $\alpha_s(m_c)\approx 0.35$ and $\alpha_s(m_b)\approx 0.21$ the
second-order correction exceeds the first-order correction even for the
bottomonium states. For charmonium, the second-order term is almost twice
as large as the already sizeable first-order correction. [Note that
the BLM estimate of the two-loop correction \cite{BC97} is far off 
the exact two-loop result.] Perturbative
matching at the scale $\mu=m_Q$ does not seem to work. Can the factorization
of short- and long-distance effects still be useful?

A novel, and perhaps unexpected, aspect at the two-loop level is the
factorization scale dependence of the non-relativistic decay constant and,
hence, the quarkonium wave function at the origin. The scale dependence
is large, especially due to the non-abelian term in Eq.~(\ref{andim}).
This scale dependence of the wave function indicates the limitation of
the non-relativistic potential model approach already at leading order in
$v^2$, since the wave functions obtained from solving the Schr\"odinger
equation are scale-independent. Despite this shortcoming it may be argued
that the wave function/non-relativistic decay constant obtained from
potential models corresponds -- if to anything -- to the wave function
evaluated at a scale typical for the bound state and not $m_Q$. This
point of view is also evident if the wave function is computed
non-perturbatively using lattice NRQCD, in which case the ultraviolet
cut-off/factorization scale is also much smaller than $m_Q$. We
consider $\mu=1\,$GeV as an adequate bound state scale for bottomonium and
charmonium, as the applicability of perturbation theory prevents us from
taking yet smaller scales. We then find, for bottomonium ($m_b=5\,$GeV, 
$n_f=4$):
\begin{equation}
\label{bottom}
f_{\Upsilon(nS)} = \left(1-\frac{8\alpha_s(m_b)}{3\pi}-1.74
\left(\frac{\alpha_s(m_b)}{\pi}\right)^2+\ldots\,\right)
f_{\Upsilon(nS)}^{\rm NR}(1\,{\mbox{GeV}}) \approx
0.81\,f_{\Upsilon(nS)}^{\rm NR}(1\,{\mbox{GeV}}).
\end{equation}
The numerical factor 0.81 is stable against variations of the scale of the
coupling at fixed factorization scale $1\,$GeV. At this factorization scale
the second-order correction is numerically insignificant. Although we do not
know whether the three-loop correction would also be small at the low
factorization scale, we tend to consider Eq.~(\ref{bottom}) as an
accurate prediction. This prediction could be tested, if
$f_{\Upsilon(nS)}^{\rm NR}(1\,{\mbox{GeV}})$ were accurately known,
for example from NRQCD lattice simulations. The large scale dependence
raises the theoretical question (the answer to which we postpone to the
long write-up) as to whether it is necessary to resum the logarithms
$\ln(m^2_Q/\mu^2)$ to all orders.

The situation is less favourable for charmonium states. Since
$m_c\approx 1.5\,$GeV, the size of the second-order correction is
altered little for scales $\mu>1\,$GeV. We have not succeeded to find
a trustworthy interpretation of Eq.~(\ref{rel}) and conclude that the
factorization of short-distance and long-distance effects may not be
useful in practice for charmonium. Since the leptonic decay is the
simplest conceivable decay, this puts into question
the possibility to obtain
universal relations between various charmonium decays and production
processes through the use of NRQCD \cite{BBL,revs}. This
pessimistic conclusion may be biased by our use of the
$\overline{\mbox{MS}}$ factorization scheme. It is conceivable that
other factorization schemes or relations between physical 
observables, from which the wave function at the 
origin is eliminated, exhibit
better convergence properties of their perturbative series. A definitive
conclusion on this issue can be obtained only once a second quarkonium 
decay is computed to two-loop order.\\

\noindent While this paper was being written,
Czarnecki and Melnikov \cite{CM} also considered the two-loop
matching of the electro-magnetic current, also using the 
asymptotic expansion method of \cite{BS}. After correction of  
a trivial normalization error (A. Czarnecki and K. Melnikov, private 
communication), their result agrees with ours.\\ 

\noindent {\em Acknowledgements.} We thank G.~Buchalla for useful
discussions and A.~Hoang for correspondence.
The work of V.S. has been supported by the Russian Foundation for
Basic Research, project 96--01--00726, and by INTAS, grant 93--0744.


\end{document}